# A Global Nucleic Acid Observatory for Biodefense and Planetary Health

The Nucleic Acid Observatory Consortium[1]

[1]Nucleic.Acid.Observatory@gmail.com

## Abstract

The spread of pandemic viruses and invasive species can be catastrophic for human societies and natural ecosystems. SARS-CoV-2 demonstrated that the speed of our response is critical, as each day of delay permitted exponential growth and dispersion of the virus. Here we propose a global Nucleic Acid Observatory (NAO) to monitor the relative frequency of everything biological through comprehensive metagenomic sequencing of waterways and wastewater. By searching for divergences from historical baseline frequencies at sites throughout the world, NAO could detect any virus or invasive organism undergoing exponential growth whose nucleic acids end up in the water, even those previously unknown to science. Continuously monitoring nucleic acid diversity would provide us with universal early warning, obviate subtle bioweapons, and generate a wealth of sequence data sufficient to transform ecology, microbiology, and conservation. We call for the immediate construction of a global NAO to defend and illuminate planetary health.

## Introduction

The SARS-CoV-2 pandemic has demonstrated our profound vulnerability to exponentially spreading viruses. Even though the first cases were estimated to have occurred in November or possibly October[1], they were not detected until December[2]. Had governments acted just a few weeks earlier, the virus may have been excluded from many more countries or even eradicated before becoming a global pandemic, as was accomplished for SARS-CoV-1 in 2003[3] and Ebola in 2014-2015[4,5]. Notably, those places that responded swiftly and aggressively to SARS-CoV-2 fared best early on, including New Zealand, Vietnam, China, South Korea, and Australia[6]. When an enemy can spread exponentially, an early detection and response requires exponentially fewer resources to achieve containment and eradication.

The same logic holds for invasive species, which devastate natural ecosystems and cause immense damage to the agriculture and forestry sectors[7]. Invasions were estimated to cost an inflation-adjusted $54 billion per year in the United States[8] and a minimum of $47-$163 billion worldwide, with costs roughly doubling every six years from 1970-2017[9]. The U.S. and China have the most to lose from future invasions[10]. Analyses of containment and eradication programs targeting 136 invasive plants[11,12] and 130 insects[13] consistently found that the larger the geographic area of the infestation, the lower the odds of success [14]. The ease of detecting the target pest was one of the most critical factors associated with eradication success.

These dynamics underscore the importance of early warning systems to contain and eradicate biological invasives. Pandemic viruses and pests share one critical feature with all other living things: their genomes are made of nucleic acids. In every region of the world, these fragments of DNA and RNA are washed into local bodies of water, where new techniques have begun to permit their reliable detection[15,16]. SARS-CoV-2 has been detected in wastewater days before the first clinical cases being reported in surrounding communities[17]; wastewater sequencing has even detected and identified variants of the virus[18–21] as well as many other viruses[22–29] and antibiotic resistance genes[30]. Laboratories using metagenomic sequencing to analyze samples of river water have successfully detected signatures of species from throughout the associated watershed[15], including terrestrial macrofauna[16], and can distinguish between closely related species[31] (Fig. 1). Even low-throughput metagenomic sequencing can detect spiked-in samples of pathogen DNA in less than 8 hours[32,33]. While most of these studies suggest that shotgun metagenomics alone has a higher limit of detection relative to quantitative PCR, there are exceptions in both environmental [34,35] and clinical samples[36,37], and it is widely agreed that performing parallel unbiased and target-specific amplification before sequencing can confer qPCR-level sensitivity to specific sequences of interest while maintaining the ability to detect previously unknown agents[38,39] or strains with mutations in commonly used RT-PCR primer binding sites[22]. Therefore, a system that combines unbiased amplification to detect previously unknown or foreign biological agents with more sensitive monitoring of particular species of concern from both wastewater and natural waterways appears to offer the best of both worlds (Fig. 2).

The case study dramatizing the importance of early detection used a much older technology: in 2013, Israel's poliovirus-specific environmental monitoring program detected a nascent outbreak in wastewater samples from the town of Rahat using plaque assays and swiftly initiated mass oral vaccination, eliminating the virus before even a single child came down with paralytic symptoms[40]. Today, we have the potential to detect *any* virus or invasive species, even novel pandemic-class agents unknown to science, through the comprehensive metagenomic sequencing of waterways. Here we propose to build a Nucleic Acid Observatory to continuously monitor the global environment for past, present, and future pandemic viruses and invasive pests.

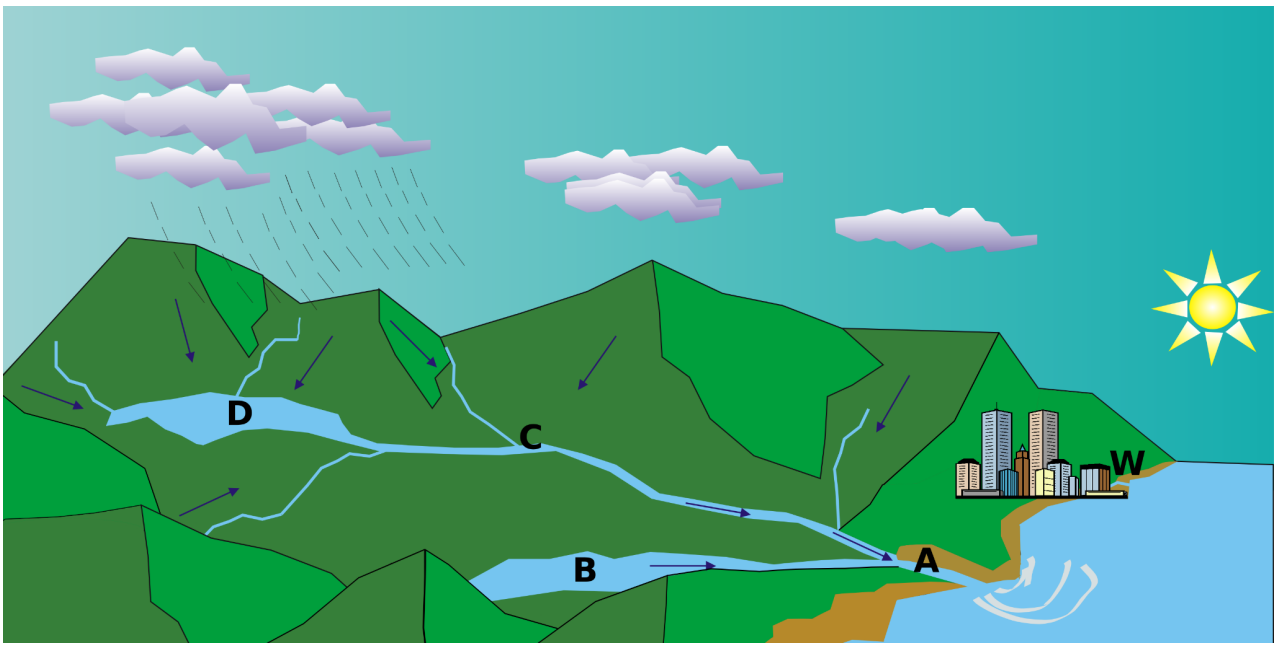

**Figure 1** | Watershed and wastewater metagenomic sequencing can detect pandemic viruses, invasive species, and engineered genes while monitoring the populations of different species. Sampling at site A detects nucleic acids from the entire watershed, including agricultural and rural regions; upstream sampling at B, C, D can pinpoint sources. Sequencing at M monitors the health of the marine ecosystem, while wastewater sequencing at W can sensitively detect human-specific pathogens.

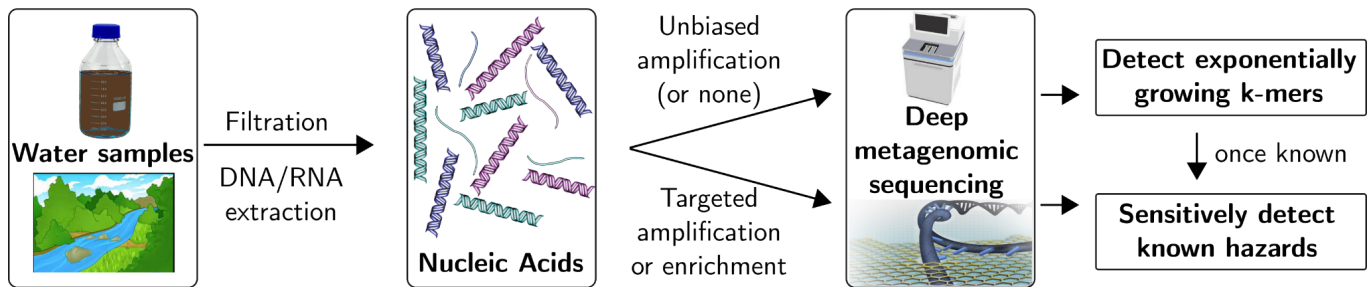

**Figure 2 |** Overview of sample processing, sequencing, and analysis. Processed samples are sequenced with and without targeted amplification of known agents. Unbiased sequencing can detect any exponential threat by monitoring the abundance of different k-mer fragments over time. The k-mers comprising any exponentially growing biological agent will increase in frequency as a group, permitting identification of the agent. Once identified, the threat can be added to the list of targets for sequence-specific amplification or enrichment, which will maximize sensitivity and immediately determine whether the new threat is present at any other sites of the Observatory. Targeted sequencing of stored samples could determine the time of introduction to any given site, aiding subsequent investigations. Alternative sequence-specific detection methods such as RT-PCR or LAMP may be used as appropriate.

## Biodefense

The world's demonstrated vulnerability to SARS-CoV-2 appears reason enough to invest in a comprehensive early-warning system. The United States has lost more citizens to the pandemic than it has in all military conflicts in the past century, yet it devotes less than 1% of its defense budget to biodefense[41]. Indeed, not a single line of the nearly $700 billion 2020 defense appropriations bill mentioned anything biological[42]. Most of the small investment in biosecurity is focused on anthrax and other chemical weapon equivalents, whose theoretical worst-case tolls are orders of magnitude lower than those of pandemics and other autonomously spreading pathogens while being far less accessible due to the need for a sophisticated aerosol delivery system[43,44].

Agriculture and natural ecosystems are equally, if not more, susceptible to autonomous pathogens. Historical fungal blights afflicting staple crops have led to massive famines in the past, most infamously in Ireland; many still cause tremendous losses today[45]. Concerted outbreaks afflicting multiple crops could be devastating for the world, especially given our increasing reliance on near-monocultures[46]. Livestock can also suffer: the ~1% case fatality rate of SARS-COV-2 is dwarfed by the 80-100% lethality of African swine fever in pigs[47]. An estimated 13-27% of all the world's swine were lost to the 2018-2019 outbreak[48].

For the moment, we largely lack the capacity to engineer biological agents capable of spreading invasively in the wild. The sole current exception, CRISPR-based gene drive technologies capable of editing populations of sexually reproducing organisms[49], underscores the critical importance of a Nucleic Acid Observatory. Relative to viruses, gene drive systems are slower to spread and far more reliably countered[50], as any given example can be overwritten by building and releasing a corresponding "immunizing reversal" drive system[49,51]. Even so, a harmful drive system could cause tremendous damage if permitted to grow exponentially; for some species, individuals with relevant technical skills could build and release such a construct single-handedly. Genome sequencing, whether of at-risk organisms or of metagenomes, appears to be the only reliable method of detecting engineered gene drive systems in the environment, and is therefore required for defense. Despite defense projects aimed at sensitively detecting engineered sequences in the environment[52], no such system has been constructed at scale.

The possibility that individuals could single-handedly edit wild species had never been previously imagined before the invention of CRISPR-based gene drive, raising the unsettling possibility that we may stumble upon similarly unanticipated ways of engineering exponential spread. Even if the unknown unknowns prove harmless, a number of emerging threats are clearly visible today. For example, several current avenues of research promise to identify publicly[53–55] and reveal ways to enhance and weaponize potentially pandemic viruses either deliberately[56] or by better understanding factors governing transmissibility[57] or immune evasion[58]. Such agents would be less readily countered than would a gene drive construct[59,60]. Constructing an early warning system could simultaneously deter attack and maximize the time available for defenders to develop and apply suitable countermeasures.

To be effective, monitoring systems should not rely on any particular phenotype, symptom profile, or sequence to be present in future natural outbreaks or bioweapons. Since very few viruses have been sequenced to date, detection systems based on known sequences may not be sensitive to future pandemic agents[61]. For example, multiplexed CRISPR-based diagnostics capable of detecting all current human pathogens have tremendous medical potential[62], but they could miss future zoonotic agents. Older systems attempting to detect aerosolized agents[63] are similarly limited to known threats, although they could be upgraded by applying metagenomic sequencing. Most importantly, any remotely sophisticated adversary would deliberately engineer a biological weapon to lack the sequences detected by any pre-existing diagnostics and defenses.

In contrast, the only way to evade detection by shotgun metagenomic sequencing would be to somehow engineer a pathogen that entirely lacks sequence-amenable nucleic acids in its genome. Since all known living organisms possess genomes that can be sequenced with current technologies and all nucleic acids present in the environment are thought to be detectable in the water[15,64], analyses searching for sequences diverging from historical baselines could sensitively detect nucleic acids from any biological construct exhibiting exponential growth. As such, a Nucleic Acid Observatory could reliably detect any and all subtle pandemic viruses, bioweapons, or other autonomous biological agents targeting humans, agriculture, or natural ecosystems within days or weeks of introduction.

## Conservation

Invasive species are a primary cause of extinction of native species around the world[65–67]. Any waterway-based metagenomic sequencing system capable of reliably detecting a pandemic virus or CRISPR-based gene drive system in a terrestrial organism living in the associated watershed should also be capable of detecting invasive species. Consider forests. The American chestnut was once the most common tree in North America; thanks to the invasive chestnut blight, it is nearly extinct. Other species have been devastated or are threatened by other agents such as Dutch elm disease, the emerald ash borer, Asian longhorned beetle, hemlock woolly adelgid, the gypsy moth, Rapid Ohi'a Death, and more[14,68,69]. Even farmed trees are vulnerable; citrus greening disease has destroyed the orange industry in Florida, costing the state 34,000 jobs[70]. Assuming that future invasions would be equivalently costly[10], the NAO could nearly pay for itself simply by enabling detection and eradication before they become widespread.

In addition to detecting invasive species, environmental DNA in rivers can provide information on the abundance of native species, including terrestrial mammals[71]. Such methods have helped scientists monitor numerous endangered species[72–75], including those that are challenging to track[76,77]. In aquatic environments, metagenomic sequencing was demonstrated to be superior to all conventional methods for species tracking[78]. A weekly census of species occupancy and abundance in environments could direct conservation resources more efficiently, and may be particularly relevant to assisting species imperiled by climate change.

A key question is whether shotgun metagenomics can monitor abundance as effectively as the metabarcoding more commonly used to detect particular marker genes in eukaryota. Recent studies suggest that the metagenomics approaches commonly used by microbial ecologists are nearly as sensitive[79]. In addition, sufficiently deep metagenomic sequencing permits the direct monitoring of genetic diversity within populations, which is widely considered a better metric of species robustness than simple abundance[80–84].

## Science

The environmental sequencing of waterways has revealed that a large fraction to an overwhelming majority of sequences of nucleic acids in the environment are not found in current repositories[27,85–87]. In other words, much of life remains entirely unknown to us.

The scientific benefits of sequencing most of the remaining genetic diversity on Earth would extend far beyond conservation; indeed, they are difficult to describe without superlatives. Microbial and macrofaunal ecology, population genetics, geochemistry, evolutionary biology, and many more disciplines would be utterly transformed by such a treasure trove of data. Virtually all of the tangible benefits arising from biological research were enabled or inspired by natural systems, from vaccines to antibiotics, aspirin to insulin, and genetic engineering, DNA sequencing, PCR and CRISPR. Sequence most living things, and we can reasonably expect to discover many more tools that may become pillars of biotechnology and medicine.

While the anticipated scientific benefits of a one-time spatial survey of genomic abundance and diversity would be staggering, the Nucleic Acid Observatory would go beyond a single snapshot by monitoring genetic diversity throughout the world for an extended time[88]. Large spatial and temporal series are considered the gold standard for ecological and evolutionary studies, but are seldom collected due to the considerable expense relative to the typically small funding streams available for ecological studies. For example, the entire U.S. National Ecological Observatory Network was funded for $434 million[89], which is a rounding error compared to the cost of COVID-19 ($5.73 trillion in relief bills passed by the U.S. Congress alone) and the $3.2-$16 trillion estimated cost of the pandemic [90,91]. If implemented in an even moderately standardized manner, the NAO would generate far and away the most comprehensive and useful ecological dataset ever collected.

## Technical options

On a high level, the Nucleic Acid Observatory would involve extracting nucleic acids from filters or concentrated water samples from rivers and sewage systems at many sites throughout the world, selectively amplifying any known sequences of concern that may be present, conducting metagenomic sequencing to generate snapshots of the nucleic acid diversity on that day[27], performing bioinformatic analyses to screen for novel sequences that have become exponentially more common relative to past snapshots, then assembling those sequences to identify the responsible organism or virus. Each of these steps can be economically and technically optimized.

### Experimental strategies

There are a variety of sampling methods that should be evaluated to determine which combinations are optimal for cost-effectiveness at different depths and breadths of threat coverage. Sample collection routinely employs artificial filters, generally developed for medical use and adapted to environmental purposes to concentrate nucleic acids[64,92–94]. More recent developments include the use of passive samplers to which eDNA will bind[95] and aquatic organisms that naturally concentrate exogenous DNA that can be cultivated and harvested, such as filter-feeding sponges, shellfish and other organisms[96], with the caveat that concentrations of bivalves may or may not substantially reduce the total concentration of eDNA[97,98] and the net dynamics of eDNA movement determinants in marine environments remain to be established[98]. Finally, the use of settlement plates and centrifugal concentration of water samples may also have utility in some settings[64]. Each approach has strengths and limitations. Artificial filters require periodic replacement and once samples are collected then storage may be challenging. Some automation of both approaches has been achieved in field deployable remote automated underwater vehicles and there is significant capacity to improve this further[99]. Living filters such as mussels and other shellfish also concentrate exogenous nucleic acids and may be suitable for monitoring, including some in pre-existing populations, but this concept remains very much in early development[96,100].

Once concentrated and extracted, nucleic acids are typically selectively or indiscriminately amplified by one of several methods[101]. For the Nucleic Acid Observatory, selective and unbiased approaches would be used in parallel to sensitively detect known sequences of interest while retaining the ability to identify previously unknown sequences that are exponentially increasing in abundance.

Different combinations of sampling, filtering, extraction, and amplification methods differ in which source organisms and sequences they are able to detect[15,18,19,102–104]. Therefore several distinct methods should be used at each testing site to ensure coverage of diverse threats. For example, three protocols may be used to target DNA from cellular organisms, viral and cell-free DNA, and viral and cell-free RNA.

Once amplified, samples may be shotgun sequenced by short or long-read sequencing, or both. Short-read sequencing based on Illumina technology is currently the most cost-effective on a per-read basis and consequently offers the greatest sensitivity, meaning it may be optimal for wastewater[105]. Nanopore sequencing offers much longer reads, superior recognition of gene drive systems and other constructs that combine elements normally never found adjacent to one another, and the ability to sequence DNA containing non-standard bases such as the newly discovered aminoadenine[106–108]. It can also sequence RNA directly, offering a means of monitoring RNA viruses without reverse transcription[109,110]. Recent advances enabling adaptive nanopore sequencing to better detect low-abundance samples may allow it to approach the sensitivity of short-read sequencing[111]. For all methods, samples from different sites can be shipped to a single central laboratory located elsewhere, barcoded, and subjected to pooled sequencing[112]. Nanopore sequencing, which offers portable sequencing in remote environments, may be required in any areas where sample stability and logistics do not permit shipment.

Sequencing costs for both short and long-read sequencing have dropped considerably faster than Moore's Law (Fig. 3), a trend that looks set to continue given a variety of early-stage alternatives to current practice[113]. Therefore, the efficacy of a NAO can reasonably be expected to grow with time.

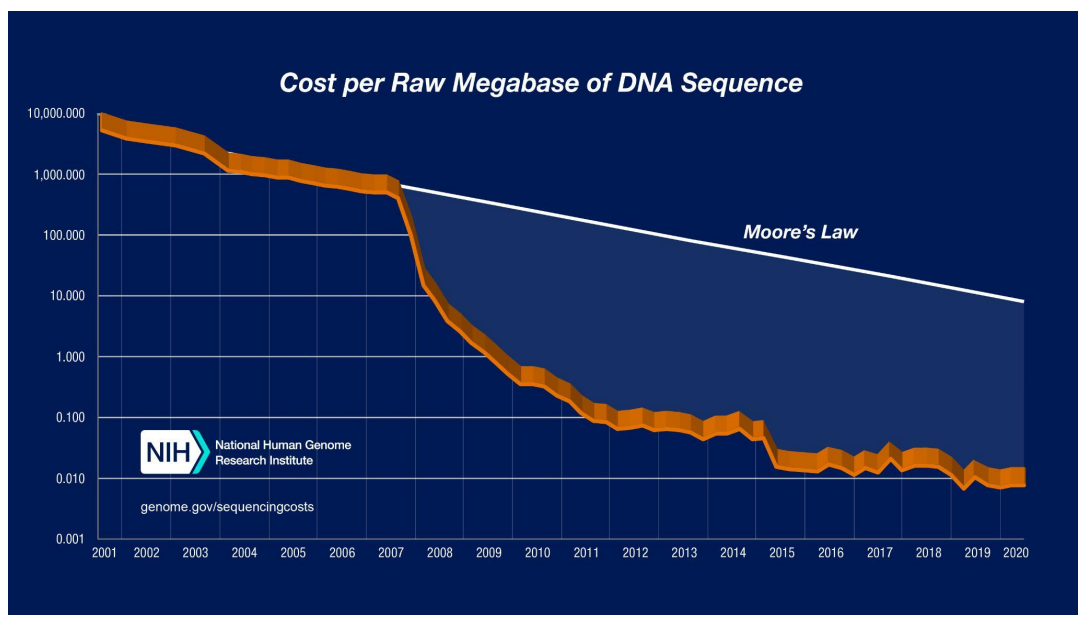

**Figure 3** | Historical cost of sequencing one million base pairs of DNA. Adopted from NHGRI[114].

## Computational strategies

Once samples are collected and sequenced, the resulting data must then be bioinformatically analyzed and interpreted. Distinct strategies are needed to monitor known versus unknown threats. Known threats—agents or genetic elements with known sequences—can be detected by mapping sequencing reads (or their protein translations) to databases of reference sequences[115–117] or by using classification, prediction, or screening methods derived from such databases[116,118]. Such databases and methods have been used to detect federally-designated select agents, human pathogens, genetically engineered elements, and bacterial genes associated with toxin production and antimicrobial resistance[30,115,116,119–121]. Once a new threat is discovered and sequenced anywhere in the world, such as when SARS-CoV-2 was first sequenced in China in January of 2020 [122], its presence can be monitored at all NAO testing sites. The number of reads mapped to a threat can also be used to quantitatively track its abundance through time by using comparisons to stable, common organisms like pepper mild mottle virus to convert counts to calibrated abundances across samples[123]. Comprehensive taxonomic profiling of specific groups or even all known species[117,118,124] can also be used to monitor for unusual deviations from a testing site's typical profile. Finally, mutational variants in a specific organism of interest can be monitored by aligning reads to its reference sequence, as recently used to track SARS-CoV-2 variants in wastewater[18,19].

However, such reference-based approaches are ill-suited to detecting a truly novel threat whose sequence is not known *a priori*. For these unknown threats, we suggest employing a reference-free strategy that looks for signatures from arbitrary sequences that have begun to exponentially increase in frequency at a given location. A signature currently used by reference-free methods for studying variation in human genomic data[125–127] and bacterial metagenomic data[128,129] are k-mers—sequences that are k base pairs long. K-mers ranging from ~30 to 40 base pairs in length are typically used because they are highly specific to the source sequence while remaining derivable directly from short sequence reads with little chance of sequencing error. An adaptation of these methods to detect increasing frequencies of sequences at a given testing location might count the occurrences of each k-mer in each new sample, then perform a statistical test to determine whether each k-mer has begun to exponentially increase in relative abundance compared to housekeeping reference genes at a recent point in time. Increasing k-mers that overlap can be assembled into longer sequences[130,131]. K-mers or assembled sequences may be matched to their containing organisms by mapping to reference databases or used to design primers for targeted amplification of the surrounding sequence[132,133]. Variations on this basic approach should be explored and may each be more or less sensitive to distinct threats.

Additional bioinformatic analyses that may prove useful for detecting emerging threats include metagenome assembly methods[117] to generate reference genomes for the "microbial dark matter" that is absent from existing databases, creating a "pan-genome" of all sequences[134] that are found under normal circumstances at a given testing site, and cloud-based pipelines for pathogen detection and taxonomic identification[135].

## Computational challenges

The NAO will generate a wealth of environmental sequencing data that will vastly supersede what is currently available. As described above, this data holds great promise for transforming ecology, microbiology, and conservation biology; but its sheer volume poses technical and operational challenges in facilitating its use. Ongoing declines in the cost of hard drives and other storage technologies mean that storing the data in "cold storage" presents only a marginal cost increase over the basal cost of sequencing (Supplementary Tables 1-4). The primary challenge will involve making the data from thousands of testing sites available to the scientific community. For this purpose, we propose storing a small, strategically chosen subset of the data in a cloud computing platform, from which users can download the data or operate on it directly by purchasing computing resources in the cloud environment[136]. A possible subsetting strategy is to store all of the most recent week's data but just weekly or monthly snapshots for later periods. The methods used to generate the data by various "collectors" (see below) may vary in space and time, making it essential to keep detailed metadata to allow researchers to account for this methodological variation in their analyses[136].

## Operations

Nations could adopt a variety of approaches to implement their own Nucleic Acid Observatory, from government-run to entirely outsourced. However, all successful operations will incentivize the reliable detection of rare nucleic acid sequences. In security, quality control is assured by consistently challenging the defenses. In this Observatory context, adequate sensitivity might be assured by employing "red teams" to simulate attacks by introducing foreign nucleic acids at various levels without informing those in charge of sample collection and sequencing.

Specifically, each monitoring site will necessarily be frequented by collectors tasked with acquiring and sequencing samples. These are members of the "blue team", whose job is to defend the area by swiftly detecting any invasive sequences. To ensure detection is sufficiently sensitive and reliable, "red team" inspectors frequently challenge the defenses by releasing known amounts of foreign DNA or RNA into the environment within each watershed or sewer system. Incentives of collectors should be tied to their success or failure to detect these foreign sequences, while inspectors should be incentivized to accurately quantify the sensitivity and identify flaws. Ideally, each site will be monitored by at least two competing organizations of sample collectors (whether private, public, or military) and be challenged by at least two competing organizations of inspectors, with incentives to discourage collusion.

The resulting competition would encourage innovation seeking greater cost-effectiveness in sample acquisition, processing, and sequencing approaches while enhancing sensitivity. For example, the establishment phase of a NAO could see different groups exploring various collection methods, sampling frequencies, and intensities to determine which approaches can offer the most sensitive and comprehensive monitoring.

## Cost, healthcare, and privacy

The cost of a Nucleic Acid Observatory will depend on the desired sensitivity for different nucleic acids and the benefits of scaling, but our back-of-the-envelope calculations estimate a total annual cost of $700 million for a pilot system monitoring wastewater from all 328 U.S. Ports of Entry and all 378 major USGS-designated water basins, in addition to one-time system setup expenses. For a complete system that would additionally monitor all major U.S. towns and cities, most international airports, and either the 378 water basins or all 2278 designated watersheds, we estimate a total annual cost of $5-15 billion annually (Supplementary Tables 1-4). Given the annual damages inflicted by invasive pests and especially by COVID-19, this looks like a remarkable bargain, one that would additionally boost employment throughout the nation (Supplementary Information Executive Summary). Other nations could enjoy similar benefits.

Sample acquisition, processing, and sequencing represent the bulk of the estimated costs, with data storage and analysis being comparatively less expensive. These costs could potentially be reduced by adapting improvements from clinical sequencing such as superior sample barcoding[112] and optimization of filter use, sample storage, and amplification techniques. Most importantly, overall costs are expected to continue to decline faster than Moore's Law with the price of sequencing (Fig. 3). As the entire global DNA sequencing market was estimated at only $8-12 billion in 2020-21, investing these sums should achieve substantial bulk efficiencies and discounts.

Accelerating the reduction of sequencing costs will offer synergistic benefits with clinical sequencing for healthcare applications. Indeed, sufficiently widespread sequencing of patient samples could plausibly substitute for wastewater sequencing in detecting human pathogens. Wastewater sequencing is somewhat more cost-effective than sequencing many clinical samples, but the primary advantage of wastewater sequencing to detect human pathogens is that it preserves privacy.

Whereas sequencing a clinical sample necessarily acquires data allowing genomic identification of the patient, the mixing of many samples in sewage precludes the identification of specific human genomes and the potential disclosure of any health, relationship, or location information. Laws and regulations requiring consent forms tightly constrain precisely what can be done with information obtained from clinical samples. In contrast, the anonymity of environmentally collected data would allow a NAO focused on wastewater and waterways to begin monitoring for any and all harmful invasives – not just those directly attacking humans – sooner rather than later.

## Conclusion

The NAO would tell us when to act when we are confronted with an emerging pandemic or other exponentially spreading biothreat. It will create a genome repository for virtually all life on Earth, give us contemporary snapshots of species health, and eventually, an historical record of ecological changes. The time to build it is now.

## References

1. Pekar, J., Worobey, M., Moshiri, N., Scheffler, K. & Wertheim, J. O. Timing the SARS-CoV-2 index case in Hubei province. *Science* (2021) doi:10.1126/science.abf8003.
2. Huang, C. *et al.* Clinical features of patients infected with 2019 novel coronavirus in Wuhan, China. *Lancet* (2020).
3. Chew, S. K. SARS: how a global epidemic was stopped. *Bull. World Health Organ.* **85**, 324 (2007).
4. Frieden, T. R. & Damon, I. K. Ebola in West Africa--CDC's Role in Epidemic Detection, Control, and Prevention. *Emerg. Infect. Dis.* **21**, 1897–1905 (2015).
5. Dahl, B. A. *et al.* CDC's Response to the 2014-2016 Ebola Epidemic - Guinea, Liberia, and Sierra Leone. *MMWR Suppl* **65**, 12–20 (2016).
6. Pearce, N., Lawlor, D. A. & Brickley, E. B. Comparisons between countries are essential for the control of COVID-19. *Int. J. Epidemiol.* **49**, 1059–1062 (2020).
7. Robinson, A. P. *Invasive Species: Risk Assessment and Management*. (Cambridge University Press, 2017).
8. Pimentel, D., Zuniga, R. & Morrison, D. Update on the environmental and economic costs associated with alien-invasive species in the United States. *Ecol. Econ.* **52**, 273–288 (2005).
9. Diagne, C. *et al.* High and rising economic costs of biological invasions worldwide. *Nature* **592**, 571–576 (2021).
10. Paini, D. R. *et al.* Global threat to agriculture from invasive species. *Proc. Natl. Acad. Sci. U. S. A.* **113**, 7575–7579 (2016).
11. Pluess, T. *et al.* Which factors affect the success or failure of eradication campaigns against alien species? *PLoS One* **7**, e48157 (2012).
12. Pluess, T. *et al.* When are eradication campaigns successful? A test of common assumptions. *Biol. Invasions* **14**, 1365–1378 (2012).
13. Tobin, P. C. *et al.* Determinants of successful arthropod eradication programs. *Biol. Invasions* **16**, 401–414 (2014).
14. Liebhold, A. M. *et al.* Eradication of Invading Insect Populations: From Concepts to Applications. *Annu. Rev. Entomol.*

**61**, 335–352 (2016).
15. Deiner, K., Fronhofer, E. A., Mächler, E., Walser, J.-C. & Altermatt, F. Environmental DNA reveals that rivers are conveyer belts of biodiversity information. *Nat. Commun.* **7**, 12544 (2016).
16. Reddington, K. *et al.* Metagenomic analysis of planktonic riverine microbial consortia using nanopore sequencing reveals insight into river microbe taxonomy and function. *Gigascience* **9**, (2020).
17. Peccia, J. *et al.* Measurement of SARS-CoV-2 RNA in wastewater tracks community infection dynamics. *Nat. Biotechnol.* **38**, 1164–1167 (2020).
18. Crits-Christoph, A. *et al.* Genome Sequencing of Sewage Detects Regionally Prevalent SARS-CoV-2 Variants. *MBio* **12**, (2021).
19. Pérez-Cataluña, A. *et al.* Detection of genomic variants of SARS-CoV-2 circulating in wastewater by high-throughput sequencing. *bioRxiv* (2021) doi:10.1101/2021.02.08.21251355.
20. Izquierdo-Lara, R. *et al.* Monitoring SARS-CoV-2 Circulation and Diversity through Community Wastewater Sequencing, the Netherlands and Belgium. *Emerg. Infect. Dis.* **27**, 1405–1415 (2021).
21. Kreier, F. The myriad ways sewage surveillance is helping fight COVID around the world. *Nature* (2021) doi:10.1038/d41586-021-01234-1.
22. Adriaenssens, E. M. *et al.* Viromic Analysis of Wastewater Input to a River Catchment Reveals a Diverse Assemblage of RNA Viruses. *mSystems* **3**, (2018).
23. Fernandez-Cassi, X. *et al.* Metagenomics for the study of viruses in urban sewage as a tool for public health surveillance. *Sci. Total Environ.* **618**, 870–880 (2018).
24. Lun, J. H., Crosbie, N. D. & White, P. A. Genetic diversity and quantification of human mastadenoviruses in wastewater from Sydney and Melbourne, Australia. *Sci. Total Environ.* **675**, 305–312 (2019).
25. Bibby, K. *et al.* Metagenomics and the development of viral water quality tools. *npj Clean Water* **2**, 1–13 (2019).
26. Kitajima, M. *et al.* SARS-CoV-2 in wastewater: State of the knowledge and research needs. *Sci. Total Environ.* **739**, 139076 (2020).
27. Nieuwenhuijse, D. F. *et al.* Setting a baseline for global urban virome surveillance in sewage. *Sci. Rep.* **10**, 13748 (2020).
28. Bisseux, M. *et al.* Monitoring of enterovirus diversity in wastewater by ultra-deep sequencing: An effective complementary tool for clinical enterovirus surveillance. *Water Res.* **169**, 115246 (2020).
29. Zahedi, A., Monis, P., Deere, D. & Ryan, U. Wastewater-based epidemiology-surveillance and early detection of waterborne pathogens with a focus on SARS-CoV-2, Cryptosporidium and Giardia. *Parasitol. Res.* (2021) doi:10.1007/s00436-020-07023-5.
30. Hendriksen, R. S. *et al.* Global monitoring of antimicrobial resistance based on metagenomics analyses of urban sewage. *Nat. Commun.* **10**, 1124 (2019).
31. Garrido-Sanz, L., Senar, M. À. & Piñol, J. Estimation of the relative abundance of species in artificial mixtures of insects using low-coverage shotgun metagenomics. *MBMG* **4**, e48281 (2020).
32. Israeli, O. *et al.* Rapid identification of unknown pathogens in environmental samples using a high-throughput sequencing-based approach. *Heliyon* **5**, e01793 (2019).
33. Li, B. Molecular and metagenomic characterization of bacterial pathogens and antibiotic resistance genes in complex environments. (University of Hawai'i at Manoa, 2020).
34. Plaire, D., Puaud, S., Marsolier-Kergoat, M.-C. & Elalouf, J.-M. Comparative analysis of the sensitivity of metagenomic sequencing and PCR to detect a biowarfare simulant (Bacillus atrophaeus) in soil samples. *PLoS One* **12**, e0177112 (2017).
35. McArdle, A. J. & Kaforou, M. Sensitivity of shotgun metagenomics to host DNA: abundance estimates depend on bioinformatic tools and contamination is the main issue. *Access Microbiol* **2**, acmi000104 (2020).
36. Miller, S. *et al.* Laboratory validation of a clinical metagenomic sequencing assay for pathogen detection in cerebrospinal fluid. *Genome Res.* **29**, 831–842 (2019).
37. van Boheemen, S. *et al.* Retrospective Validation of a Metagenomic Sequencing Protocol for Combined Detection of RNA and DNA Viruses Using Respiratory Samples from Pediatric Patients. *J. Mol. Diagn.* **22**, 196–207 (2020).
38. Chiu, C. Y. & Miller, S. A. Clinical metagenomics. *Nat. Rev. Genet.* **20**, 341–355 (2019).
39. Deng, X. *et al.* Metagenomic sequencing with spiked primer enrichment for viral diagnostics and genomic surveillance. *Nat Microbiol* **5**, 443–454 (2020).

40. Brouwer, A. F. *et al.* Epidemiology of the silent polio outbreak in Rahat, Israel, based on modeling of environmental surveillance data. *Proc. Natl. Acad. Sci. U. S. A.* **115**, E10625–E10633 (2018).
41. Sell, T. K. & Watson, M. Federal agency biodefense funding, FY2013-FY2014. *Biosecur. Bioterror.* **11**, 196–216 (2013).
42. U.S. Senate Appropriations Committee. United States Department of Defense Appropriations, 2020. (2019).
43. Riedel, S. Anthrax: a continuing concern in the era of bioterrorism. *Proc.* **18**, 234–243 (2005).
44. Morens, D. M. & Fauci, A. S. Emerging Pandemic Diseases: How We Got to COVID-19. *Cell* **182**, 1077–1092 (2020).
45. Haas, B. J. *et al.* Genome sequence and analysis of the Irish potato famine pathogen Phytophthora infestans. *Nature* **461**, 393–398 (2009).
46. Brooks, D. R., Hoberg, E. P., Boeger, W. A. & Trivellone, V. Emerging infectious disease: An underappreciated area of strategic concern for food security. *Transbound. Emerg. Dis.* (2021) doi:10.1111/tbed.14009.
47. Schulz, K., Conraths, F. J., Blome, S., Staubach, C. & Sauter-Louis, C. African Swine Fever: Fast and Furious or Slow and Steady? *Viruses* **11**, (2019).
48. Weaver, T. R. D. & Habib, N. Evaluating Losses Associated with African Swine Fever in the People's Republic of China and Neighboring Countries. *Asian Development Bank* (2020).
49. Esvelt, K. M., Smidler, A. L., Catteruccia, F. & Church, G. M. Concerning RNA-guided gene drives for the alteration of wild populations. *Elife* e03401 (2014).
50. Esvelt, K. Gene drive: the thing to fear is fear itself. (2018).
51. DiCarlo, J. E., Chavez, A., Dietz, S. L., Esvelt, K. M. & Church, G. M. Safeguarding CRISPR-Cas9 gene drives in yeast. *Nat. Biotechnol.* **33**, 1250–1255 (2015).
52. Finding Engineering-Linked Indicators (FELIX). https://www.iarpa.gov/index.php?option=com_content&view=article&id=999&Itemid=397.
53. Tumpey, T. M. *et al.* Characterization of the reconstructed 1918 Spanish influenza pandemic virus. *Science* **310**, 77–80 (2005).
54. Carroll, D. *et al.* The Global Virome Project. *Science* **359**, 872–874 (2018).
55. Grange, Z. *et al.* SpillOver: A new tool for ranking the risk of viral spillover to humans using big data. *Front. Vet. Sci.* **6**, (2019).
56. Herfst, S. *et al.* Airborne transmission of influenza A/H5N1 virus between ferrets. *Science* **336**, 1534–1541 (2012).
57. Davies, N. G. *et al.* Estimated transmissibility and impact of SARS-CoV-2 lineage B.1.1.7 in England. *Science* (2021) doi:10.1126/science.abg3055.
58. Beachboard, D. C. & Horner, S. M. Innate immune evasion strategies of DNA and RNA viruses. *Curr. Opin. Microbiol.* **32**, 113–119 (2016).
59. Esvelt, K. M. Inoculating science against potential pandemics and information hazards. *PLoS Pathog.* **14**, e1007286 (2018).
60. Lewis, G., Millett, P., Sandberg, A., Snyder-Beattie, A. & Gronvall, G. Information Hazards in Biotechnology. *Risk Anal.* **39**, 975–981 (2019).
61. Wille, M., Geoghegan, J. L. & Holmes, E. C. How accurately can we assess zoonotic risk? *PLoS Biol.* **19**, e3001135 (2021).
62. Ackerman, C. M. *et al.* Massively multiplexed nucleic acid detection with Cas13. *Nature* **582**, 277–282 (2020).
63. Institute of Medicine (US), National Research Council (US) Committee on Effectiveness of National Biosurveillance Systems: Biowatch & the Public Health System. *The BioWatch System*. (National Academies Press (US), 2011).
64. Bowers, H. A. *et al.* Towards the Optimization of eDNA/eRNA Sampling Technologies for Marine Biosecurity Surveillance. *Water* **13**, 1113 (2021).
65. Spatz, D. R. *et al.* Globally threatened vertebrates on islands with invasive species. *Sci Adv* **3**, e1603080 (2017).
66. Jones, H. P. *et al.* Invasive mammal eradication on islands results in substantial conservation gains. *Proceedings of the National Academy of Sciences* **113**, 4033–4038 (2016).
67. Dueñas, M.-A., Hemming, D. J., Roberts, A. & Diaz-Soltero, H. The threat of invasive species to IUCN-listed critically endangered species: A systematic review. *Global Ecology and Conservation* **26**, e01476 (2021).
68. Brockerhoff, E. G., Liebhold, A. M., Richardson, B. & Suckling, D. M. Eradication of invasive forest insects: concepts, methods, costs and benefits. *New Zealand Journal of Forestry Science. 40 suppl. : S117-S135.* **40**, S117–S135 (2010).
69. Holmes, T. P., Aukema, J. E., Von Holle, B., Liebhold, A. & Sills, E. Economic impacts of invasive species in forest past,

present, and future. *In: The Year In Ecology and Conservation Biology, 2009. Ann. NY Acad. Sci. 1162: 18-38.* **1162**, 18–38 (2009).
70. Singerman, A. & Useche, M. P. Impact of citrus greening on citrus operations in Florida. *EDIS* **2016**, 4–4 (2020).
71. Sales, N. G. *et al.* Fishing for mammals: Landscape-level monitoring of terrestrial and semi-aquatic communities using eDNA from riverine systems. *J. Appl. Ecol.* **57**, 707–716 (2020).
72. Rees, H. C., Baker, C. A., Gardner, D. S., Maddison, B. C. & Gough, K. C. The detection of great crested newts year round via environmental DNA analysis. *BMC Res. Notes* **10**, 327 (2017).
73. Loeza-Quintana, T. *et al.* Environmental DNA detection of endangered and invasive species in Kejimkujik National Park and Historic Site. *Genome* **64**, 172–180 (2021).
74. Farrell, J. A., Whitmore, L. & Duffy, D. J. The Promise and Pitfalls of Environmental DNA and RNA Approaches for the Monitoring of Human and Animal Pathogens from Aquatic Sources. *Bioscience* doi:10.1093/biosci/biab027.
75. Adams, C. I. M. *et al.* Beyond Biodiversity: Can Environmental DNA (eDNA) Cut It as a Population Genetics Tool? *Genes* **10**, (2019).
76. Bakker, J. *et al.* Environmental DNA reveals tropical shark diversity in contrasting levels of anthropogenic impact. *Sci. Rep.* **7**, 16886 (2017).
77. Lafferty, K. D., Benesh, K. C., Mahon, A. R., Jerde, C. L. & Lowe, C. G. Detecting Southern California's White Sharks With Environmental DNA. *Frontiers in Marine Science* **5**, 355 (2018).
78. McElroy, M. E. *et al.* Calibrating Environmental DNA Metabarcoding to Conventional Surveys for Measuring Fish Species Richness. *Frontiers in Ecology and Evolution* **8**, 276 (2020).
79. Singer, G. A. C., Shekarriz, S., McCarthy, A., Fahner, N. & Hajibabaei, M. The utility of a metagenomics approach for marine biomonitoring. *Cold Spring Harbor Laboratory* 2020.03.16.993667 (2020) doi:10.1101/2020.03.16.993667.
80. Saccheri, I. *et al.* Inbreeding and extinction in a butterfly metapopulation. *Nature* **392**, 491–494 (1998).
81. Gibson, G. & Dworkin, I. Uncovering cryptic genetic variation. *Nat. Rev. Genet.* **5**, 681–690 (2004).
82. Hayden, E. J., Ferrada, E. & Wagner, A. Cryptic genetic variation promotes rapid evolutionary adaptation in an RNA enzyme. *Nature* **474**, 92–95 (2011).
83. Weeks, A. R., Stoklosa, J. & Hoffmann, A. A. Conservation of genetic uniqueness of populations may increase extinction likelihood of endangered species: the case of Australian mammals. *Front. Zool.* **13**, 31 (2016).
84. Ralls, K. *et al.* Call for a paradigm shift in the genetic management of fragmented populations. *Conserv. Lett.* **11**, e12412 (2018).
85. Tamaki, H. *et al.* Metagenomic analysis of DNA viruses in a wastewater treatment plant in tropical climate. *Environ. Microbiol.* **14**, 441–452 (2012).
86. Alhamlan, F. S., Ederer, M. M., Brown, C. J., Coats, E. R. & Crawford, R. L. Metagenomics-based analysis of viral communities in dairy lagoon wastewater. *J. Microbiol. Methods* **92**, 183–188 (2013).
87. Ng, T. F. F. *et al.* High variety of known and new RNA and DNA viruses of diverse origins in untreated sewage. *J. Virol.* **86**, 12161–12175 (2012).
88. Davies, N. *et al.* A call for an international network of genomic observatories (GOs). *Gigascience* **1**, 5 (2012).
89. Pennisi, E. A groundbreaking observatory to monitor the environment. *Science* **328**, 418–420 (2010).
90. Walmsley, T., Rose, A. & Wei, D. The Impacts of the Coronavirus on the Economy of the United States. *Econ Disaster Clim Chang* 1–52 (2020).
91. Cutler, D. M. & Summers, L. H. The COVID-19 Pandemic and the $16 Trillion Virus. *JAMA* **324**, 1495–1496 (2020).
92. Fulton-Bennett, K. An autonomous vehicle coupled with a robotic laboratory proves its worth. https://www.mbari.org/lrauv-and-esp/ (2019).
93. Water Quality Monitoring and Sampling. https://aqualytical.com/.
94. Microbiology Society. Ocean Robots Uncover Microbial Secrets. https://microbiologysociety.org/publication/past-issues/oceans/article/ocean-robots-uncover-microbial-secrets.html.
95. Bessey, C. *et al.* Passive eDNA collection enhances aquatic biodiversity analysis. *Commun Biol* **4**, 236 (2021).
96. Mariani, S., Baillie, C., Colosimo, G. & Riesgo, A. Sponges as natural environmental DNA samplers. *Curr. Biol.* **29**, R401–R402 (2019).
97. Friebertshauser, R., Shollenberger, K., Janosik, A., Garner, J. T. & Johnston, C. The effect of bivalve filtration on eDNA-based detection of aquatic organisms. *PLoS One* **14**, e0222830 (2019).

98. Shogren, A. J., Tank, J. L., Egan, S. P., Bolster, D. & Riis, T. Riverine distribution of mussel environmental DNA reflects a balance among density, transport, and removal processes. *Freshw. Biol.* **64**, 1467–1479 (2019).
99. Yamahara, K. M. *et al.* In situ Autonomous Acquisition and Preservation of Marine Environmental DNA Using an Autonomous Underwater Vehicle. *Frontiers in Marine Science* **6**, 373 (2019).
100. Mariani, S. *et al.* Estuarine molecular bycatch as a landscape-wide biomonitoring tool. *bioRxiv* 2021.01.10.426097 (2021) doi:10.1101/2021.01.10.426097.
101. Taberlet, P., Bonin, A., Zinger, L. & Coissac, E. *Environmental DNA: For Biodiversity Research and Monitoring*. (Oxford Univ Pr, 2018).
102. Deiner, K., Walser, J.-C., Mächler, E. & Altermatt, F. Choice of capture and extraction methods affect detection of freshwater biodiversity from environmental DNA. *Biol. Conserv.* **183**, 53–63 (2015).
103. McLaren, M. R., Willis, A. D. & Callahan, B. J. Consistent and correctable bias in metagenomic sequencing experiments. *Elife* **8**, (2019).
104. Fitzpatrick, A. H. *et al.* High Throughput Sequencing for the Detection and Characterization of RNA Viruses. *Front. Microbiol.* **12**, 621719 (2021).
105. Foox, J. *et al.* Multi-Platform Assessment of DNA Sequencing Performance using Human and Bacterial Reference Genomes in the ABRF Next-Generation Sequencing Study. *bioRxiv* 2020.07.23.218602 (2020) doi:10.1101/2020.07.23.218602.
106. Zhou, Y. *et al.* A widespread pathway for substitution of adenine by diaminopurine in phage genomes. *Science* **372**, 512–516 (2021).
107. Pezo, V. *et al.* Noncanonical DNA polymerization by aminoadenine-based siphoviruses. *Science* **372**, 520–524 (2021).
108. Sleiman, D. *et al.* A third purine biosynthetic pathway encoded by aminoadenine-based viral DNA genomes. *Science* **372**, 516–520 (2021).
109. Grädel, C. *et al.* Whole-Genome Sequencing of Human Enteroviruses from Clinical Samples by Nanopore Direct RNA Sequencing. *Viruses* **12**, (2020).
110. Wongsurawat, T. *et al.* Rapid Sequencing of Multiple RNA Viruses in Their Native Form. *Front. Microbiol.* **10**, 260 (2019).
111. Martin, S. *et al.* Nanopore adaptive sampling: a tool for enrichment of low abundance species in metagenomic samples. *bioRxiv* 2021.05.07.443191 (2021) doi:10.1101/2021.05.07.443191.
112. Dai, M. *et al.* One-Seq: A Highly Scalable Sequencing-Based Diagnostic for SARS-CoV-2 and Other Single-Stranded Viruses. *medRxiv* 2021.04.12.21253357 (2021).
113. The Cost of Sequencing a Human Genome. https://www.genome.gov/about-genomics/fact-sheets/Sequencing-Human-Genome-cost.
114. DNA Sequencing Costs: Data. https://www.genome.gov/about-genomics/fact-sheets/DNA-Sequencing-Costs-Data.
115. Gu, W., Miller, S. & Chiu, C. Y. Clinical Metagenomic Next-Generation Sequencing for Pathogen Detection. *Annu. Rev. Pathol.* **14**, 319–338 (2019).
116. Albin, D. *et al.* SeqScreen: a biocuration platform for robust taxonomic and biological process characterization of nucleic acid sequences of interest. in *2019 IEEE International Conference on Bioinformatics and Biomedicine (BIBM)* 1729–1736 (2019).
117. Breitwieser, F. P., Lu, J. & Salzberg, S. L. A review of methods and databases for metagenomic classification and assembly. *Brief. Bioinform.* **20**, 1125–1136 (2019).
118. Ondov, B. D. *et al.* Mash Screen: high-throughput sequence containment estimation for genome discovery. *Genome Biol.* **20**, 232 (2019).
119. Chaturvedi, R., Brenner, P., Martineau, M., Posard, M. N. & Merrill, M. *Assessing the Need for and Uses of Sequences of Interest Databases: A Report on the Proceedings of a Two-day Workshop*. (RAND Corporation, 2019).
120. Sichtig, H. *et al.* FDA-ARGOS is a database with public quality-controlled reference genomes for diagnostic use and regulatory science. *Nat. Commun.* **10**, 3313 (2019).
121. Wang, Q., Kille, B., Liu, T. R., Elworth, R. A. L. & Treangen, T. J. PlasmidHawk improves lab of origin prediction of engineered plasmids using sequence alignment. *Nat. Commun.* **12**, 1167 (2021).
122. Novel 2019 coronavirus genome. https://virological.org/t/novel-2019-coronavirus-genome/319 (2020).
123. Wu, F. *et al.* SARS-CoV-2 Titers in Wastewater Are Higher than Expected from Clinically Confirmed Cases. *mSystems*

124. Kim, D., Song, L., Breitwieser, F. P. & Salzberg, S. L. Centrifuge: rapid and sensitive classification of metagenomic sequences. *Genome Res.* **26**, 1721–1729 (2016).
125. Chong, Z. *et al.* novoBreak: local assembly for breakpoint detection in cancer genomes. *Nat. Methods* **14**, 65–67 (2017).
126. Rahman, A., Hallgrímsdóttir, I., Eisen, M. & Pachter, L. Association mapping from sequencing reads using k-mers. *Elife* **7**, (2018).
127. Standage, D. S., Brown, C. T. & Hormozdiari, F. Kevlar: A Mapping-Free Framework for Accurate Discovery of De Novo Variants. *iScience* **18**, 28–36 (2019).
128. Wang, Y. *et al.* Identifying Group-Specific Sequences for Microbial Communities Using Long k-mer Sequence Signatures. *Front. Microbiol.* **9**, 872 (2018).
129. Wang, Y., Chen, Q., Deng, C., Zheng, Y. & Sun, F. KmerGO: A Tool to Identify Group-Specific Sequences With k-mers. *Front. Microbiol.* **11**, 2067 (2020).
130. Huang, X. & Madan, A. CAP3: A DNA sequence assembly program. *Genome Res.* **9**, 868–877 (1999).
131. Simpson, J. T. *et al.* ABySS: a parallel assembler for short read sequence data. *Genome Res.* **19**, 1117–1123 (2009).
132. Boutigny, A.-L., Fioriti, F. & Rolland, M. Targeted MinION sequencing of transgenes. *Sci. Rep.* **10**, 15144 (2020).
133. Alquezar-Planas, D. E. *et al.* DNA sonication inverse PCR for genome scale analysis of uncharacterized flanking sequences. *Methods Ecol. Evol.* **12**, 182–195 (2021).
134. Wang, Q., Elworth, R. A. L., Liu, T. R. & Treangen, T. J. Faster pan-genome construction for efficient differentiation of naturally occurring and engineered plasmids with plaster. in (Schloss Dagstuhl - Leibniz-Zentrum fuer Informatik GmbH, Wadern/Saarbruecken, Germany, 2019). doi:10.4230/LIPICS.WABI.2019.19.
135. Kalantar, K. L. *et al.* IDseq-An open source cloud-based pipeline and analysis service for metagenomic pathogen detection and monitoring. *Gigascience* **9**, (2020).
136. Langmead, B. & Nellore, A. Cloud computing for genomic data analysis and collaboration. *Nat. Rev. Genet.* **19**, 208–219 (2018).

# Authors

The Nucleic Acid Observatory Consortium includes many contributing authors who will be listed in the peer-reviewed publication. The following contributors were tasked with preprint preparation and coordination. They are listed here for administrative and procedural purposes only:

Michael R. McLaren
Department of Population Health and Pathobiology, North Carolina State University, Raleigh, NC, USA

Adam H. Marblestone
Federation of American Scientists, USA

Kevin M. Esvelt
Media Lab, Massachusetts Institute of Technology, Cambridge, MA, USA

The authors declare no conflicts of interest.

# Supplementary Tables

| Assumed parameters | | Notes |
|---|---:|---|
| Sequencing machine cost | $600,000 | Will shrink with future technologies |
| Sequencing run cost | $1,500 | Reagents are highly profitable for manufacturers. Current cost for NovaSeq at a sequencing core is ~$4k; this number assumes a large scaling discount and normal yearly cost reduction by the time of implementation. |
| Reads per machine run | 2.00E+09 | Parameter of sequencer. |
| Bits per machine run | 4.00E+11 | Parameter of sequencer |
| Days per machine run | 1.2 | Parameter of sequencer, and of read length |
| Reads needed per sample | 1.00E+09 | This review; can tradeoff with number of samples per site to some extent |
| Samples needed per site per day | 3 | Airport wastewater is lower volume, therefore fewer reads needed than a city; assume 1/5 |
| Sites needed in U.S. | 328 | |
| Total yearly salary per site | $100,000 | 1 person per site at $100k/year/person on average |
| Data storage cost per terabyte | $10 | Reasonable by the time this is available |
| Rent per year per 10 sites | $500,000 | Assume labs are in expensive cities, but some consolidation through shipping is possible. 365 large packages overnight << $500k per year. |
| Filters and sample prep costs per sample | $20 | Uncertain of scaling; rough estimate |
| | | |
| **Derived parameters** | | |
| Machines needed per site | 1.80E+00 | |
| Machines needed total | 5.90E+02 | |
| Investment for waterway tech development | $100,000,000 | |
| *Up front cost* | $454,240,000 | |
| Yearly sequencing reagent/run cost total | $269,370,000 | |
| Yearly filters or other sample prep cost total | $7,183,200 | |
| Yearly salary total | $32,800,000 | |
| Yearly data storage total | $89,790 | |
| Yearly rent total | $16,400,000 | |
| *Ongoing yearly running cost estimate* | $325,842,990 | |

**Supplementary Table 1 |** Rough estimated cost of a pilot Nucleic Acid Observatory to sequence wastewater from all 328 Ports of Entry in the United States. Assumes use of Illumina Novaseq or equivalent. Costs will decline as sequencing technology continues to improve.

| Assumed parameters | | Notes |
|---|---|---|
| Sequencing machine cost | $600,000 | Will shrink with future technologies |
| Sequencing run cost | $1,500 | Reagents are highly profitable for manufacturers. Current cost for NovaSeq at a sequencing core is ~$4k; this number assumes a large scaling discount and normal yearly cost reduction by the time of implementation. |
| Reads per machine run | 2.00E+09 | Parameter of sequencer. |
| Bits per machine run | 4.00E+11 | Parameter of sequencer |
| Days per machine run | 1.2 | Parameter of sequencer, and of read length |
| Reads needed per sample | 1.00E+09 | This review; can tradeoff with number of samples per site to some extent |
| Samples needed per site per day | 3 | Airport wastewater is lower volume, therefore fewer reads needed than a city; assume 1/5 |
| Sites needed in U.S. | 706 | 328 Ports of Entry + 378 USGS water basins |
| Total yearly salary per site | $100,000 | 1 person per site at $100k/year/person on average |
| Data storage cost per terabyte | $10 | Reasonable by the time this is available |
| Rent per year per 10 sites | $500,000 | Assume labs are in expensive cities, but some consolidation through shipping is possible. 365 large packages overnight << $500k per year. |
| Filters and sample prep costs per sample | $20 | Uncertain of scaling; rough estimate |
| | | |
| **Derived parameters** | | |
| Machines needed per site | 1.80E+00 | |
| Machines needed total | 1.27E+03 | |
| *Up front sequencing machine cost* | $762,480,000 | |
| Yearly sequencing reagent/run cost total | $579,802,500 | |
| Yearly filters or other sample prep cost total | $15,461,400 | |
| Yearly salary total | $70,600,000 | |
| Yearly data storage total | $193,268 | |
| Yearly rent total | $35,300,000 | |
| *Ongoing yearly running cost estimate* | $701,357,168 | |

**Supplementary Table 2 |** Rough estimated cost of a pilot Nucleic Acid Observatory to sequence wastewater from all 328 Ports of Entry in the United States as well as water from the 378 USGS-designated water basins. Costs will decline as sequencing technology continues to improve.

| Assumed parameters | | Notes |
|---|---:|---|
| Sequencing machine cost | $600,000 | Will shrink with future technologies |
| Sequencing run cost | $900 | Reagents are highly profitable for manufacturers. Current cost for NovaSeq at a sequencing core is ~$4k; this number assumes a large scaling discount and normal yearly cost reduction by the time of implementation. |
| Reads per machine run | 2.00E+09 | Parameter of sequencer. |
| Bits per machine run | 4.00E+11 | Parameter of sequencer |
| Days per machine run | 1.2 | Parameter of sequencer, and of read length |
| Reads needed per sample | 1.00E+09 | This review; can tradeoff with number of samples per site to some extent |
| Samples needed per site per day | 10 | Smaller sites will require fewer, big cities/rivers need more; assume this is an average. Also trades off with number of reads per sample. |
| Sites needed in world | 1470 | |
| Total yearly salary per site | $250,000 | 1 person per site at $100k/year/person on average |
| Data storage cost per terabyte | $10 | Reasonable by the time this is available |
| Rent per year per 10 sites | $500,000 | Assume labs are in expensive cities, but some consolidation through shipping is possible. 365 large packages overnight << $500k per year. |
| Filters and sample prep costs per sample | $20 | Uncertain of scaling; rough estimate |
| | | |
| **Derived parameters** | | |
| Machines needed per site | 6.00E+00 | |
| Machines needed total | 8.82E+03 | |
| *Up front sequencing machine cost* | $5,292,000,000 | |
| Yearly sequencing reagent/run cost total | $2,414,475,000 | |
| Yearly filters or other sample prep cost total | $107,310,000 | |
| Yearly salary total | $367,500,000 | |
| Yearly data storage total | $1,341,375 | |
| Yearly rent total | $73,500,000 | |
| *Ongoing yearly running cost estimate* | $2,964,126,375 | |

**Supplementary Table 3 |** Rough estimated cost of a Nucleic Acid Observatory to sequence wastewater from all 328 U.S. Ports of Entry, water from the 378 USGS-designated water basins, wastewater from the 314 cities over 100,000 people, and wastewater from the 450 international airports with flights to the United States. Initial costs could be reduced by performing exclusively targeted sequencing at some days and sites, and will decline as sequencing technology continues to improve.

| Assumed parameters | | Notes |
|---|---:|---|
| Sequencing machine cost | $600,000 | Will shrink with future technologies |
| Sequencing run cost | $900 | Reagents are highly profitable for manufacturers. Current cost for NovaSeq at a sequencing core is ~$4k; this number assumes a large scaling discount and normal yearly cost reduction by the time of implementation. |
| Reads per machine run | 2.00E+09 | Parameter of sequencer. |
| Bits per machine run | 4.00E+11 | Parameter of sequencer |
| Days per machine run | 1.2 | Parameter of sequencer, and of read length |
| Reads needed per sample | 1.00E+09 | This review; can tradeoff with number of samples per site to some extent |
| Samples needed per site per day | 10 | Smaller sites will require fewer, big cities/rivers need more; assume this is an average. Also trades off with number of reads per sample. |
| Sites needed in world | 5013 | |
| Total yearly salary per site | $250,000 | 1 person per site at $100k/year/person on average |
| Data storage cost per terabyte | $10 | Reasonable by the time this is available |
| Rent per year per 10 sites | $500,000 | Assume labs are in expensive cities, but some consolidation through shipping is possible. 365 large packages overnight << $500k per year. |
| Filters and sample prep costs per sample | $20 | Uncertain of scaling; rough estimate |
| | | |
| **Derived parameters** | | |
| Machines needed per site | 6.00E+00 | |
| Machines needed total | 3.01E+04 | |
| *Up front sequencing machine cost* | $18,046,800,000 | |
| Yearly sequencing reagent/run cost total | $8,233,852,500 | |
| Yearly filters or other sample prep cost total | $365,949,000 | |
| Yearly salary total | $1,253,250,000 | |
| Yearly data storage total | $4,574,363 | |
| Yearly rent total | $250,650,000 | |
| *Ongoing yearly running cost estimate* | $10,108,275,863 | About the estimated size of the sequencing industry if current trends continue until ~2026 |

**Supplementary Table 4 |** Rough estimated cost of a Nucleic Acid Observatory to sequence wastewater from all 328 U.S. Ports of Entry, water from all 2264 USGS-designated watersheds, and wastewater from the 1521 towns and cities with over 25,000 people as well as the 900 major international airports. Initial costs could be reduced by performing exclusively targeted sequencing at some days and sites, and will decline as sequencing technology continues to improve.

# Nucleic Acid Observatory

Executive Summary

The United States lost over 600,000 Americans and trillions of dollars to COVID-19.

- Early detection is critical to controlling exponentially growing biological threats
- Future natural and engineered threats to humans and ecosystems are highly likely
- Intensive environmental DNA sequencing can detect all such threats
- Sequencing wastewater and natural waterways does not endanger genetic privacy

**Phase I**: Pilot study on vulnerable sites (immediate)

- $700m per year for 3 years, +$760m initial setup, that will sequence:
    - daily wastewater samples from all 328 CBP Ports of Entry ($325m annually + $350m setup)
    - biweekly samples from all 378 USGS water basins ($375m annually + $410m setup)

**Phase II**: Nucleic Acid Observatory (beginning in 2023)

- $3.0b per year, +$5.3b initial setup, for deeper sequencing of daily samples from:
    - all 328 Ports of Entry
    - all 378 water basins
    - all 314 U.S. cities over 100,000 people
    - 450 international airports with flights to the United States
- $10.4b per yr, +$18.4b setup, for deeper sequencing of daily samples from:
    - all 328 Ports of Entry
    - all 2264 watersheds
    - all 1521 U.S. towns over 25,000 people
    - 900 international airports

Side benefits for agriculture, science, and the environment:

- Invasive pests costing billions per year can be detected early enough for eradication
- Deep sequencing will unearth new molecules for the biotech industry
- Sequencing waterways can monitor the abundance of all species

Sensitivity should more than double every two years:

- DNA sequencing costs have fallen a billion-fold over twenty years
- The rate of improvement recently slowed to 'only' the level of Moore's Law
- At current rates, the sensitivity of the system will double each year
- NAO would double the size of the sequencing market, catalyzing further acceleration

Defends against pandemics and biological weapons engineered to be subtle:

- All living things can be detected with sufficiently deep sequencing
- Searching for sequences that swiftly become more common can detect all threats